\documentclass[iop]{emulateapj}

\newcommand{\myemail}{gruane@astro.caltech.edu}

\slugcomment{Accepted: June 22, 2017}

\shorttitle{Deep search for planets forming around TW Hya}
\shortauthors{Ruane et al.}

\begin{document}


\title{Deep imaging search for planets forming in the TW Hya protoplanetary disk with the Keck/NIRC2 vortex coronagraph}


\author{G.~Ruane\altaffilmark{1,2}, D.~Mawet\altaffilmark{1,3}, J.~Kastner\altaffilmark{4,5}, T.~Meshkat\altaffilmark{3,6}, M.~Bottom\altaffilmark{3}, B.~Femen\'{i}a Castell\'{a}\altaffilmark{7}, O.~Absil\altaffilmark{8,9}, C.~Gomez~Gonzalez\altaffilmark{8}, E.~Huby\altaffilmark{8,10}, Z.~Zhu\altaffilmark{11}, R.~Jensen-Clem\altaffilmark{1}, \'{E}. Choquet\altaffilmark{3,12}, E. Serabyn\altaffilmark{3}}

\email{\myemail}
\altaffiltext{1}{Department of Astronomy, California Institute of Technology, Pasadena, CA 91125, USA}
\altaffiltext{2}{NSF Astronomy and Astrophysics Postdoctoral Fellow}
\altaffiltext{3}{Jet Propulsion Laboratory, California Institute of Technology, Pasadena, CA 91109, USA}
\altaffiltext{4}{School of Physics \& Astronomy, Laboratory for Multiwavelength Astrophysics, Rochester Institute of Technology, Rochester, NY 14623, USA}
\altaffiltext{5}{Chester F. Carlson Center for Imaging Science, Rochester Institute of Technology, Rochester, NY 14623, USA}
\altaffiltext{6}{IPAC, California Institute of Technology, Pasadena, CA 91125, USA}
\altaffiltext{7}{W. M. Keck Observatory, Kamuela, HI 96743, USA}
\altaffiltext{8}{Space Sciences, Technologies, and Astrophysics Research (STAR) Institute, Universit\'{e} de Li\`{e}ge, Li\`{e}ge, Belgium}
\altaffiltext{9}{F.R.S.-FNRS Research Associate}
\altaffiltext{10}{LESIA, Observatoire de Paris, Meudon, France}
\altaffiltext{11}{Department of Physics and Astronomy, University of Nevada, Las Vegas, Las Vegas, NV 89154, USA}
\altaffiltext{12}{Hubble Fellow}

\begin{abstract}
Distinct gap features in the nearest protoplanetary disk, TW Hya (distance of 59.5$\pm$0.9 pc), may be signposts of ongoing planet formation. We performed long-exposure thermal infrared coronagraphic imaging observations to search for accreting planets especially within dust gaps previously detected in scattered light and submm-wave thermal emission. Three nights of observations with the Keck/NIRC2 vortex coronagraph in $L^\prime$ (3.4-4.1~$\mu$m) did not reveal any statistically significant point sources. We thereby set strict upper limits on the masses of non-accreting planets. In the four most prominent disk gaps at 24, 41, 47, and 88 au, we obtain upper mass limits of 1.6-2.3, 1.1-1.6, 1.1-1.5, and 1.0-1.2 Jupiter masses~($M_J$) assuming an age range of 7-10~Myr for TW Hya. These limits correspond to the contrast at 95\% completeness (true positive fraction of 0.95) with a 1\% chance of a false positive within $1^{\prime\prime}$ of the star. We also approximate an upper limit on the product of planet mass and planetary accretion rate of $M_p\dot{M}\lesssim10^{-8} M_J^2/yr$ implying that any putative $\sim0.1~M_J$ planet, which could be responsible for opening the 24 au gap, is presently accreting at rates insufficient to build up a Jupiter mass within TW Hya's pre-main sequence lifetime.
\end{abstract}


\section{Introduction}

Interactions between planets and the circumstellar material from which they form manifest as large-scale dust disk density structures that appear as gaps, rings, or spirals \citep{Bryden1999,Pinilla2012,Zhu2014,Dong2015a,Dong2017,Canovas2017}. Such features have been detected in the disks around numerous nearby, young stars via sub-mm interferometry \citep{Hughes2007,Andrews2009,Isella2010,Andrews2016,vanderPlas2017} and scattered light imaging \citep{Garufi2013,Rapson2014,Benisty2015,Thalmann2015,Akiyama2015,Rapson2015}. In fact, observers need to look no further than TW Hydrae, the nearest protoplanetary disk \citep[59.5$\pm$0.9 pc,][]{Gaia2016} to find gap features that could indicate the presence of planets potentially undergoing formation. 

TW Hya is a pre-main sequence, classical T Tauri star with a mass in the range of 0.7-0.8~$M_{\odot}$ \citep{Andrews2012,Herczeg2014} and a massive circumstellar disk \citep{Bergin2013}. In addition to being the nearest young solar analog orbited by a gas-rich disk, TW Hya is one of the most promising systems for directly observing signposts of planet formation owing to its relatively advanced age for a protoplanetary disk (7-10 Myr, see section \ref{sec:age}), close to face-on geometry \citep[$\sim7^\circ$ inclination,][]{Qi2004}, and distinct radial gap features seen in scattered light \citep{Weinberger2002,Akiyama2015,Rapson2015,Debes2016_TWHyainnerstructure,Debes2017,vanBoekel2017} and thermal emission \citep{Andrews2016} from dust in the disk.

\begin{table*}
\begin{center}
\caption{TW Hya coronagraphic deep field observations\label{tab:observations}}
\begin{tabular}{ll*{6}{c}}
\tableline
\tableline
Date & Target & Frames & Integration time & DIMM seeing & Airmass & Transparency & PA rotation\\
\tableline
2017 Jan. 9 & TYC 7128-1252-1 & 78 & 59 min &$ 0\farcs 55$ &1.7-2.0 &clear & 32$^\circ$  \\
 & TW Hya & 120 & 90 min &$0\farcs 70$ &1.7-2.2 &clear &$44^\circ$\\
  & CD-31 10139 & 20 & 15 min &$0\farcs 64$ &1.6-1.9 &clear & 6$^\circ$\\

2017 Jan. 10 & TYC 7128-1252-1 & 70 & 53 min &$0\farcs 35$ &1.7-2.0 &clear & $27^\circ$\\
 & TW Hya & 120 & 90 min &$0\farcs 50$ &1.7-2.2 &clear &$45^\circ$\\
  & CD-31 10139 & 70 & 53 min &$0\farcs 55$ &1.6-1.9 &clear & $29^\circ$\\

2017 Jan. 13 & TYC 7128-1252-1 & 13 & 9.8 min &$0\farcs 40$ &1.8-2.0 &clear & $4^\circ$\\
 & TW Hya & 117 & 88 min &$0\farcs 48$ &1.7-2.2 &clear &$45^\circ$\\
  & CD-31 10139 & 78 & 59 min &$0\farcs 55$ &1.6-1.9 &clear & $30^\circ$\\
\tableline
\end{tabular}
\end{center}
\end{table*}

Observations by \citet{Andrews2016} with the Atacama Large Millimeter/submillimeter Array (ALMA) at 870$\mu$m show gaps at separations of $\sim$0\farcs02 (1~au), $\sim$0\farcs4 (24~au), $\sim$0\farcs7 (41~au), and $\sim$0\farcs8 (47~au). In addition, optical/near-infrared scattered light observations with VLT/SPHERE confirm the $\sim$24~au gap as well as a clearing at $\sim$1\farcs5, or 88~au \citep{vanBoekel2017}. 

Dynamical simulations by \citet{Dong2017} suggest planets with masses of 0.05-0.5 and 0.03-0.3 Jupiter masses ($M_J$), respectively, may be sculpting the gaps at $\sim$0\farcs4 (24~au) and $\sim$1\farcs5 (88~au). A $0.1~M_J$ protoplanet actively accreting material from the disk at rates on the order of $10^{-7} M_J/yr$, i.e. sufficient to build a Jupiter mass within TW Hya's pre-main sequence lifetime, could have an absolute magnitude in $L^\prime$ (3.4-4.1 $\mu$m) of $\sim$13, which is detectable via a high contrast imager at $\Delta L^\prime\approx10$ \citep{Zhu2015}. Moreover, such a protoplanet can only be feasibly detected than in $L^\prime$ band, or at longer infrared wavelengths (e.g. $M$ or $N$). The protoplanet would be at least 11 and 6 magnitudes fainter at $H$ and $K$ bands than in $L^\prime$ band, respectively, which is beyond the detection capability of state-of-the-art high contrast imagers at small angular separations \citep[see e.g.][]{vanBoekel2017}. 

We have performed long-exposure $L^\prime$ observations of TW~Hya using the NIRC2 vortex coronagraph at the W.M.~Keck Observatory to search for protoplanets forming within the disk. Although the images did not reveal any statistically significant point sources, our contrast sensitivity translates into strict upper limits on the masses of non-accreting planets as well as constraints on the mass accretion rates of potential protoplanets. 

\section{Age of TW Hya}\label{sec:age}

Constraints on the masses and mass accretion luminosities of young exoplanets orbiting within gaps in the TW Hya circumstellar disk obtained from direct thermal imaging depend sensitively on the assumed age of the star itself. Specifically, the younger the host star, the higher the expected luminosity of a young planet of a given mass. In the case of TW Hya, published age estimates range from $\sim$3~Myr to $\sim$10~Myr. The younger age estimates are derived from the star's inferred effective temperature and measured luminosity, which allows TW Hya to be placed relative to isochrones gleaned from pre-main sequence evolutionary models \citep{VaccaSandell2011,Donaldson2016}. The older end of the age range is based on statistical analyses of the ensemble of young stars within $\sim$15 pc of TW Hya that are evidently comoving and, hence, presumably coeval with TW Hya \citep[i.e., the TW Hya Association; e.g., ][and references therein]{Ducourant2014,Bell2015,Donaldson2016}. It is becoming increasingly apparent that the former (isochronal) age determination method, which depends on an accurate assessment of stellar effective temperature as well as the availability of robust models of the structure and atmospheres of late-type stars, may underestimate the ages of individual stars \citep[see discussions in, e.g.,][]{Kastner2015,Pecaut2016,Jeffries2017}. For that reason, and because we seek to err on the side of conservative exoplanet mass constraints from our images, we adopt an age range of 7-10 Myr for TW Hya in the analysis described in this paper.

\section{Observations and Processing}

The Keck/NIRC2 vortex coronagraph \citep{Serabyn2017,Mawet2017} is an instrument mode that enables infrared high-contrast imaging in $L^\prime$ and $M$ bands at small angular separations from the star ($\gtrsim$100~mas). This capability provides unique opportunities to detect self-luminous planets and protoplanets. While the vortex coronagraph suppresses starlight without significantly impeding the transmission of off-axis sources, angular differential imaging (ADI) and reference star differential imaging (RDI) are crucial observing strategies for optimizing the detection limits of the high contrast observations \citep{Marois2006,Lafreniere2009}. In practice, imperfect correction of atmospheric turbulence and optical aberrations cause unwanted starlight to leak through the coronagraph. The ADI strategy estimates the stellar contribution in an image from a sequence of frames with relative parallactic angle (PA) rotation and typically achieves the best sensitivity to point sources outside of a few diffracted beamwidths from the star. However, the best detection limits at small separations ($\lesssim$0\farcs3) may be obtained by RDI, which estimates the stellar contribution solely from images of similar stars \citep[see e.g.][]{Serabyn2017}.

We observed TW Hya with NIRC2 over three nights in Jan. 2017 (Table \ref{tab:observations}) under stable seeing conditions of 0\farcs57$\pm$0\farcs28, with angular resolution of $\sim$0\farcs08 and a plate scale of 0\farcs01 per pixel\footnote{https://www2.keck.hawaii.edu/inst/nirc2/genspecs.html}. Each night consisted of a $\sim$90~min integration on TW Hya, which provides the maximum PA rotation possible from Maunakea: $\sim45^\circ$. In addition, two point spread function (PSF) reference stars TYC 7128-1252-1 and CD-31 1013 were imaged directly before and after TW Hya with 10-60 min integration times. The reference stars were chosen to optimally reproduce the PSF during the TW Hya observations by matching the telescope elevation (hence declination), signal on the wavefront sensor ($R_\mathrm{mag}$), and $L^\prime$ magnitude using the WISE W1 channel as a proxy (Table \ref{tab:stars}), while also avoiding objects with infrared excess or known companions. Images with the star position offset from the focal plane mask were obtained for photometric reference and to determine the PSF morphology. The total integration time on TW Hya was 4.5 hours and an additional 4 hours for reference PSF stars targets.

After correcting for bad pixels, flat-fielding, subtracting sky background frames, and co-registering the images, we applied principal component analysis \citep[PCA;][]{Soummer2012} to estimate and subtract the stellar contribution from the images using the Vortex Image Processing (VIP) software package\footnote{https://github.com/vortex-exoplanet} \citep{VIP2015,GomezGonzalez2017}. 

\section{Statistically robust detection limits}
After subtracting off the stellar contribution, the distribution of noise in the image is approximately Gaussian \citep[see e.g.][]{Mawet2014}. Under this assumption, we chose the detection threshold for point sources such that there was a 1\% chance of having a false positive within $1^{\prime\prime}$ of the host star, which roughly corresponds to the radius of the disk in scattered light \citep{vanBoekel2017}. 
%We divided the $1^{\prime\prime}$ disk into 12.5 annuli, whose width matches the full-width half-maximum (FWHM) of the off-axis PSF, each of them having a 0.08\% chance of a false positive.

To ensure an equal probability of a false positive at all locations in the image, the false positive fraction (FPF) decreases as a function of angular separation as the number of independent and identically distributed samples within an annulus about the star increases with radius \citep{JensenClem2018}. The resulting FPF is $10^{-4}$ at the inner working angle ($\sim0\farcs1$) and $10^{-5}$ at $1^{\prime\prime}$. However, the FPF may be higher than the Gaussian model predicts in reality since the true noise distribution is unknown. We nonetheless translate the FPF into a detection threshold by inverting the cumulative distribution function of the noise assuming a Student-t distribution to account for the small number of samples \citep{Mawet2014} as described in the appendix. The resulting detection threshold varies from $8.1\;\sigma(R=1)$ at the inner working angle to $4.5\;\sigma(R=12.5)$ at $1^{\prime\prime}$, where $\sigma(R)$ is the contrast corresponding to one standard deviation in the reduced image and $R$ is the radial coordinate normalized by the full-width half-maximum (FWHM) of the off-axis PSF ($\sim$8 pixels or $\sim0\farcs08$). We used fake companion injection and retrieval to determine $\sigma(R)$ accounting for degradation of the planet signal induced by the PCA starlight subtraction algorithm  \citep[see e.g.][]{Absil2013}.

\begin{table}
\begin{center}
\caption{Star properties \label{tab:stars}}
\begin{tabular}{*{5}{c}}
\tableline
\tableline
Name & RA & DEC & $R_\mathrm{mag}^a$ & W1$^b$ \\
\tableline
TYC 7128-1252-1 & 08 01 58.3 & -33 51 36.9 & 10.23 & 6.83\\
TW Hya & 11 01 51.9 & -34 42 17.0 & 10.43 & 7.01 \\
CD-31 10139 & 13 11 29.8 & -32 29 15.1 & 10.35 & 6.60 \\
\tableline
\end{tabular}
\tablenotetext{1}{UCAC4 catalogue \citep{Zacharias2013}}
\tablenotetext{2}{WISE catalogue \citep{Wright2010}, W1 band: 3.4$\mu$m}
\end{center}
\end{table}

\begin{figure}
\centering
\includegraphics[width=\linewidth]{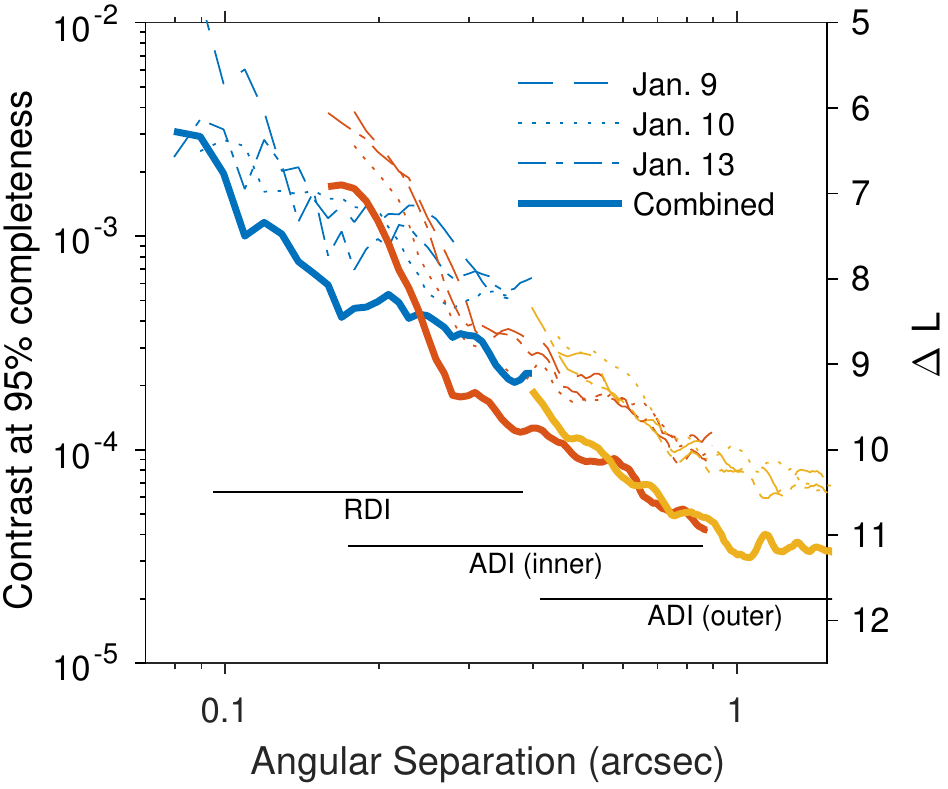}
\caption{Contrast at 95\% completeness (TPF), with a 1\% chance of a false positive within $1^{\prime\prime}$ of the star, using RDI for the smallest separations and ADI outside of $\sim$250 mas. ADI was applied in two annular regions separately: ``inner" and ``outer." The dashed and dotted lines represent the contrast limits achieved on each individual night, whereas the thicker solid lines show the performance after combining all three nights of data. \label{fig:detlimits}}
\end{figure}

The completeness, or sensitivity, of an observation is described by the true positive fraction \citep[TPF; see e.g.][]{Lafreniere2007,Wahhaj2013}. Whereas the detection threshold corresponds the 50\% completeness contour (TPF=0.5), Fig. \ref{fig:detlimits} shows the contrast at 95\% completeness (TPF=0.95), which varies from $10.1\;\sigma(R=1)$ at the inner working angle to $6.1\;\sigma(R=12.5)$ at $1^{\prime\prime}$ (see Appendix for a detailed derivation of the contrast limits).

Since RDI provides gains over ADI at small separations, we applied PCA-RDI in an annulus $0\farcs08$-$0\farcs5$ and PCA-ADI in the two remaining two annuli: an inner one over the range $0\farcs16$-$1^{\prime\prime}$ and an outer one over the range $0\farcs4$-$1\farcs75$. 

We calculated the principal components (PCs) of the combined data using all of the frames from all three nights. The optimal number of PCs (i.e. the number than minimizes $\sigma$) was 77, 27, 25 for the RDI, inner ADI, and outer ADI reductions. Since RDI does not suffer from self-subtraction effects, the optimum number of PCs is much higher than for ADI. On the other hand, both the PCA-ADI and PCA-RDI reduction schemes suffer from over-subtraction effects, which are accounted for in the fake companion injection and retrieval process.

Combining the three nights of data provided $\sim$2 times better contrast than the average of the individual nights, with both ADI and RDI approaches. We therefore conjecture that the contrast gains from combining $n$ epochs provides better than $\sqrt{n}$ improvement that one would expect with a purely Gaussian noise distribution. These gains arise from improvements in the stellar PSF model subtracted from every image provided by the inclusion of additional frames from other nights. This emphasizes the shortcomings of the typical noise assumptions applied in the interpretation of high-contrast images and planning of observations. 

\begin{figure}
\centering
\includegraphics[width=\linewidth]{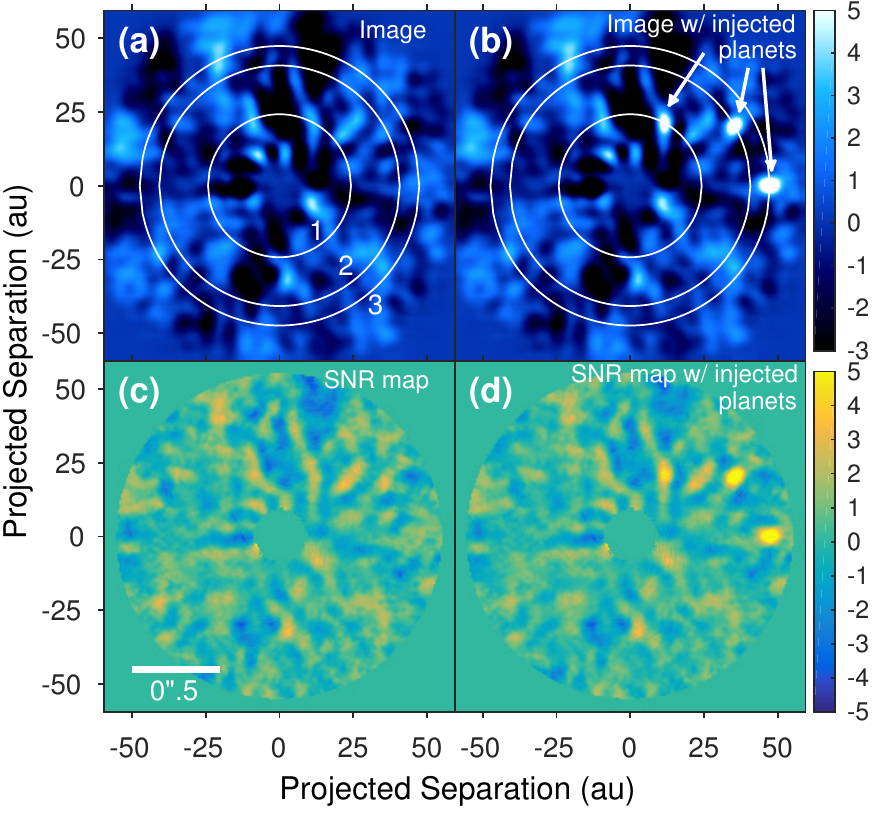}
\caption{Result of the inner ADI reduction of the combined data from all three observing nights. (a)-(b)~Images (a)~without and (b)~with injected fake companions at $\Delta L$=10.5 (or $\sim 1.5 M_J$ according to the AMES-Cond model, see section \ref{sec:mass}) at the positions of the gaps, shown by white circles. The color axis is in arbitrary units. (c)-(d)~Corresponding signal-to-noise ratio (SNR) maps confirming that (c)~no statistically significant point sources appear within 50 au of the star and (d)~the injected companions would be detected if present in the data. \label{fig:images}}
\end{figure}

Figure \ref{fig:images}a shows the residuals within the $0\farcs16$-$1^{\prime\prime}$ annulus after subtracting the reconstructed stellar image using 27 PCs. The residuals are matched-filtered using an image of the off-axis PSF. Annotations indicate the locations of the known submm continuum emission gaps at $\sim$0\farcs4 (24~au), $\sim$0\farcs7 (41~au), and $\sim$0\farcs8 (47~au), referred to as gaps 1-3, respectively \citep{Andrews2016}. The corresponding signal-to-noise ratio (SNR) map, i.e. the brightness relative to the standard deviation of independent samples in an annulus about the star (see Fig. \ref{fig:images}c), confirms that there are no statistically significant point sources after the starlight has been subtracted. However, injected fake companions at $\Delta L^\prime = 10.5$ would be clearly detected inside the gaps in the TW Hya disk (see Fig. \ref{fig:images}b,d). A point source of this brightness could correspond to a non-accreting planet of $\sim1.5~M_J$ (according to the \citet{Baraffe2003} models) or a 0.1~$M_J$ planet accreting at a rate of $\sim10^{-7}M_J/yr$ \citep{Zhu2015}. The models used to calculate the photometry of these planets are discussed in Sections \ref{sec:mass} and \ref{sec:accretion}.

\section{Upper limits on non-accreting planet mass}
\label{sec:mass}

We calculated the luminosity of a non-accreting planet using AMES-Cond, BT-Settl, and AMES-Dusty models \citep{Baraffe2003,Allard2012}. Dusty and Cond span a range from maximal to minimal dust content, whereas Settl accounts for dust formation via a paramater-free cloud model \citep[also see discussion in][]{Bowler2016}. Each provides the absolute magnitude of an exoplanet as a function of planet age and mass, which we have interpolated from precomputed grids\footnote{http://perso.ens-lyon.fr/france.allard/}.  

\begin{figure}[t]
\centering
\includegraphics[width=\linewidth]{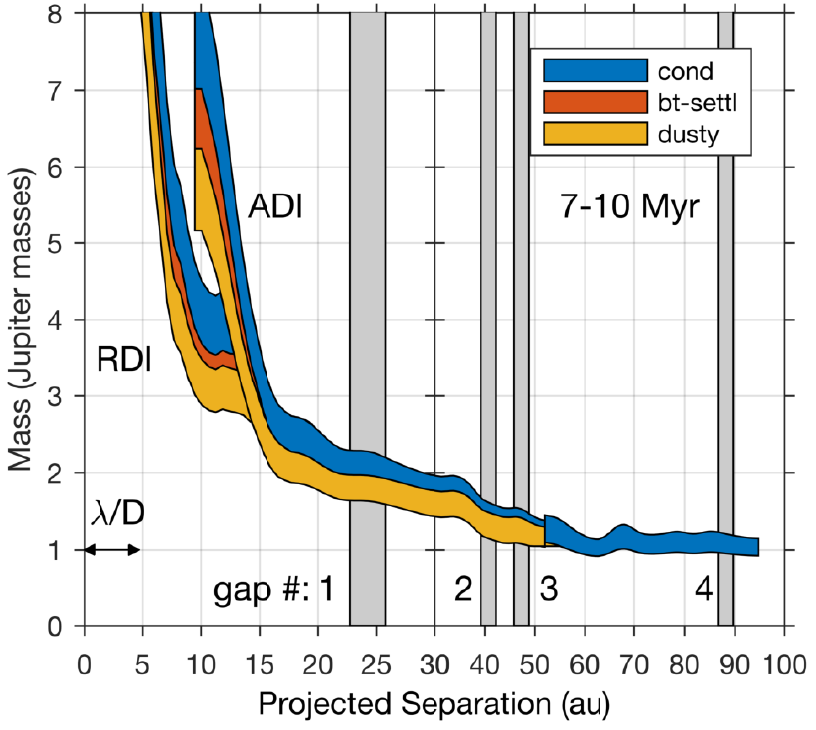}
\caption{Mass of non-accreting planetary companions corresponding to 95\% completeness (see contrast limits plotted in Fig.~\ref{fig:detlimits}) for AMES-Cond, BT-Settl, and AMES-Dusty models. The spread in the mass limits is due to the assumed range of stellar ages: 7-10~Myr. The angular resolution of the telescope (i.e. $\lambda/D$) corresponds to $\sim$5~au. \label{fig:mass}}
\end{figure}

The upper limits on the mass of non-accreting planets correspond to the contrast at 95\% completeness (see Fig.~\ref{fig:mass}). The range of masses reflects the assumed ages of 7-10 Myr. Here, we have also included the position of a fourth gap at $\sim$1\farcs5 (88~au) apparent in scattered light \citep{vanBoekel2017}, which we denote ``gap 4". For the oldest assumed age, 10 Myr, the Cond model predicts the highest masses: 2.3, 1.6, 1.5, and 1.2 $M_J$ in gaps 1-4, respectively. Dusty, on the other hand, is generally the most optimistic model in this case; assuming an age of 7 Myr implies respective planet masses of 1.6, 1.1, 1.1, and 1.0~$M_J$ in the gaps. 

A planet in the TW Hya disk is likely still undergoing formation and may therefore have an accreting circumplanetary disk, which is not included in the Cond, Settl, and Dusty models. Thus, the results above should be interpreted as upper limits for the mass of planets forming in the disk. The thermal emission from the circumplanetary disk could be much brighter than the emission from the planet owing to contraction alone. In the next section, we determine the brightness of accreting protoplanets and constrain their mass accretion rates. 

\section{Constraints on actively accreting protoplanets}
\label{sec:accretion}

\begin{figure}
\centering
\includegraphics[width=\linewidth]{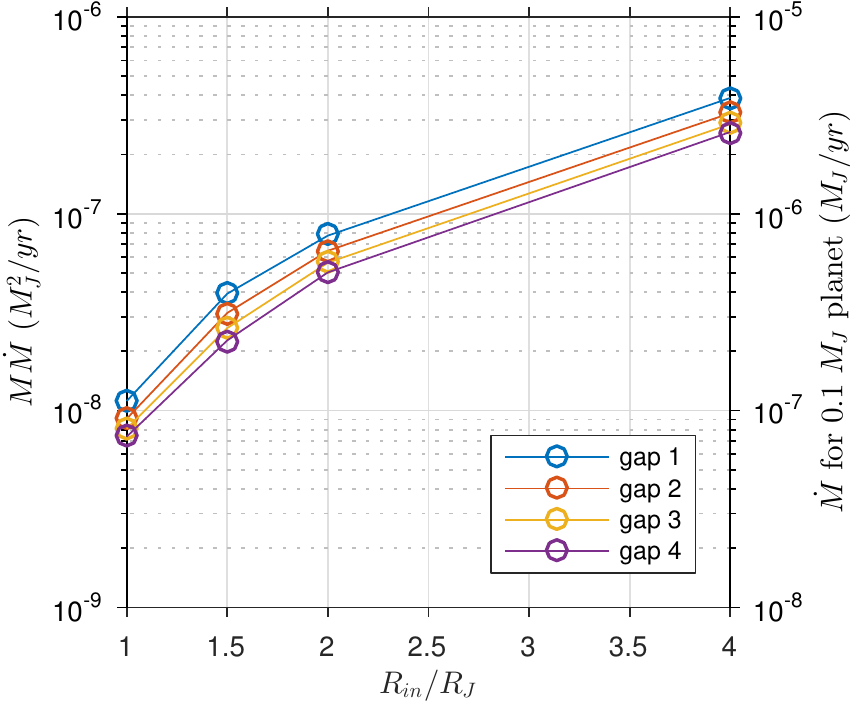}
\caption{Mass accretion limits corresponding to the contrast limits within the most prominent gaps in the TW Hya disk, given as the product of the planet mass and the mass accretion rate $M_p\dot{M}$ as well as $\dot{M}$ for a 0.1~$M_J$ planet. $R_\mathrm{in}$ is the inner radius of the circumplanetary disk and $R_J$ is the radius of Jupiter. \label{fig:limits}}
\end{figure}

\citet{Zhu2015} calculated the absolute magnitude of an accreting circumplanetary disk in near-infrared bands $J$ through $N$ and found that the brightness depends on the product of the planet mass and the mass accretion rate $M_p\dot{M}$ as well as the inner radius of the circumplanetary disk $R_\mathrm{in}$. Figure \ref{fig:limits} shows the upper limits for the accretion rate of a protoplanet forming within gaps 1-4 as a function of $R_\mathrm{in}$, as calculated from the \citet{Zhu2015} models. Owing to the inherent degeneracy between $M_p\dot{M}$ and $R_\mathrm{in}$, we are not able to place unambiguous upper limits on the accretion rate of the planet with infrared photometry alone. However, based on models of planet-disk dynamical interactions, \citet{Dong2017} estimated the mass of the planet carving out the gaps in the TW Hya disk and found that, for instance, planets with masses of 0.05-0.5 and 0.03-0.3 $M_J$, respectively, may be sculpting the gaps at $\sim$0\farcs4 (24~au) and $\sim$1\farcs5 (88~au). The \citet{Zhu2015} model predicts that a 0.1~$M_J$ protoplanet could be bright enough in $L^\prime$ to fall within the detection limits of our observations. For example, the contrast limits suggest that a planet of mass 0.1~$M_J$ accreting from a circumplanetary disk of inner radius $R_\mathrm{in}=R_J$, where $R_J$ is the radius of Jupiter, would have to be accreting at a rate $\dot{M}\lesssim10^{-7} M_J/yr$. Assuming a constant accretion rate, such a putative 0.1~$M_J$ planet would be $<$1~Myr old or must have a larger circumplanetary disk inner radius. The deep detection limits achieved in these observations imply that a planet with $R_\mathrm{in}=R_J$ is presently accreting at rates insufficient to form a Jupiter mass planet within TW~Hya's estimated lifetime of 10~Myr.

The lack of knowledge regarding $R_\mathrm{in}$ precludes a definitive upper limit for the accretion rate within the disk gaps for all protoplanets, but nonetheless our results confirm that planets in a runaway accretion phase ($M_p\dot{M}\gtrsim10^{-8}M_J^2/yr$) could be detected at infrared wavelengths $>3\mu m$ with currently available high contrast imaging instruments.

\section{Discussion \& Conclusions}

We have presented deep coronagraphic observations of TW Hya with the NIRC2 vortex coronagraph at W. M. Keck Observatory in $L^\prime$ (3.4-4.1~$\mu$m), a wavelength regime that provides unique sensitivity to self-luminous, young planets. Gaps previously detected in scattered light and thermal emission from the disk provide tantalizing evidence that planets may be sculpting features in the dust. 

Our rigorous statistical analysis resulted in strict upper limits on the mass of non-accreting planets within the disk on the order of $\sim$1-2.5~$M_J$. We also predict that actively accreting planets with much lower masses may also fall within the detection limits of these observations. However, degeneracies in the modelled brightness of circumplanetary disks do not allow us to set upper limits on both the mass of protoplanets and their accretion rates, but only the product of the planet mass and the mass accretion rate $M_p\dot{M}$ with a given disk inner radius $R_\mathrm{in}$. For $R_\mathrm{in}=R_J$, we obtain an approximate upper limit of $M_p\dot{M}\lesssim10^{-8} M_J^2/yr$ implying that a putative $\sim0.1~M_J$ mass planet, such as might be responsible for opening the 24 au gap, is presently accreting at rates insufficient to build up a Jupiter mass within TW Hya's pre-main sequence lifetime.

This work also demonstrates an optimized high-contrast imaging observing strategy that combines the benefits of ADI with RDI to enhance the contrast limits achieved at small angular separations ($\lesssim$0\farcs3), unlocking the small inner working angle performance of vortex coronagraphs, which provide the high optical throughput (down to $\sim1~\lambda/D$) needed to detect faint companions.

Although these observations are motivated by the possibility that the TW Hya gaps are induced by planet-disk interactions, there are alternate mechanisms that may cause ring structures to appear in protoplanetary disks, including condensation fronts \citep{Cuzzi2004} and zonal flows \citep{Johansen2009}. Searching for protoplanets embedded in circumstellar disks tests the dynamical hypothesis; that is, detecting a point source in a disk gap would constitute compelling evidence that planets are opening the gaps. However, our non-detection does not necessarily rule out any of these scenarios. 

Tighter constraints on the masses and accretion rates of protoplanets in the disk would require more multi-epoch coronagraphic observations with current ground-based infrared high contrast imagers equipped with small inner working angle coronagraphs in $L^\prime$ band or longer wavelengths \citep{Absil2016}. Future instrumentation may enable higher sensitivity. For example, JWST-NIRCam will provide deep contrast limits of point sources in the outer gaps of TW Hya, but the inner working angle is not small enough to peer into the 24~au gap. On the other hand, infrared adaptive optics instruments on future 30-40~m class telescopes will likely surpass the sensitivity of NIRC2, and access even lower mass planets and/or lower accretion rates.

%Circumstellar disks containing gaps and cavities are prime targets for directly witnessing the birth of giant planets, not only to add to the relatively small sample of directly imaged exoplanets, but also to provide important constraints on planet formation mechanisms essential to understanding when, where, and how planets form.
%The remarkable sensitivity to young planets at small angular separations from the star demonstrated here has inspired a survey of additional circumstellar disks containing gaps, cavities, and spirals to search for planets or protoplanets whose interaction with the disk may be generating such features. These systems are prime targets for directly witnessing the birth of giant planets, not only to add to the relatively small sample of directly imaged exoplanets, but also to provide important constraints on planet formation mechanisms essential to understanding when, where, and how planets form.

\acknowledgments
G.R. is supported by an NSF Astronomy and Astrophysics Postdoctoral Fellowship under award AST-1602444. The data presented herein were obtained at the W.M. Keck Observatory, which is operated as a scientific partnership among the California Institute of Technology, the University of California and the National Aeronautics and Space Administration (NASA). The Observatory was made possible by the generous financial support of the W.M. Keck Foundation. The authors wish to recognize and acknowledge the very significant cultural role and reverence that the summit of Maunakea has always had within the indigenous Hawaiian community.  We are most fortunate to have the opportunity to conduct observations from this mountain. J.H.K. acknowledges support from NASA Exoplanets program grant NNX16AB43G to RIT. Part of this work was carried out at the Jet Propulsion Laboratory (JPL), California Institute of Technology, under contract with NASA. The research leading to these results has received funding from the European Research Council under the European Union's Seventh Framework Programme (ERC Grant Agreement n. 337569), and from the French Community of Belgium through an ARC grant for Concerted Research Action. E.C. acknowledges support from NASA through Hubble Fellowship grant HF2-51355 awarded by STScI, which is operated by AURA, Inc. for NASA under contract NAS5-26555, for research carried out at the Jet Propulsion Laboratory, California Institute of Technology.

Facilities: \facility{W. M. Keck Observatory}, \facility{Keck:II (NIRC2)}

\clearpage

\appendix

\section{Derivation of contrast limits}

The detection limits reported here provide a fixed number of false positives per radial position $R$ in the image, where the noise distribution associated with each position is calculated in an annulus whose inner and outer radius have a mean of $R$ and difference that corresponds to one FWHM of the off-axis PSF. The false positive fraction (FPF) as a function of $R$ is given by 
\begin{equation}
\mathrm{FPF}(R)=\frac{N_\mathrm{FP}/R_\mathrm{max}}{2\pi R},
\end{equation}
where $N_{FP}$ is the acceptable number of false positives within radial distance $R<R_\mathrm{max}$. $R$ and $R_\mathrm{max}$ are normalized by the FWHM of the off-axis PSF ($\sim$8 pixels) such that $2\pi R$ is the number of independent and identically distributed (i.i.d.) samples in an annulus about the star. Here, $N_\mathrm{FP}=0.01$ and $R_\mathrm{max}=12.5$ (equivalent to $\sim1^{\prime\prime}$). The FPF in this case varies from $10^{-4}$ at the inner working angle ($\sim0\farcs1$) to $10^{-5}$ at $1^{\prime\prime}$. 

The threshold for detection as a function of radial position $\tau(R)$ is given by
\begin{equation}
\frac{\tau(R)}{\sigma(R)}=C_\mathrm{st}^{-1}\left(\left(1-\mathrm{FPF}(R)\right)|\left(2\pi R - 2\right)\right),
\end{equation}
where $\sigma(R)$ is the contrast corresponding to one standard deviation and $C_\mathrm{st}^{-1}(.)$~is the inverse of the Student-t cumulative distribution function:
\begin{equation}
C_\mathrm{st}(x|\nu)=\int_{-\infty}^x \frac{\Gamma((\nu+1)/2)}{\sqrt{\pi\nu}\Gamma(\nu/2)} \left( 1 + \frac{t^2}{\nu}\right)^{-\frac{\nu+1}{2}} dt,
\end{equation}
where $\nu$ is the number of degrees of freedom, i.e. one less than the number of independent samples. Although the noise is assumed to be normally distributed, the Student-t distribution is used to account for the small number of i.i.d. samples available within the annuli at small angular separations \citep[see discussion in][]{Mawet2014}. Specifically, the number of i.i.d. samples is $2\pi R - 1$ excluding the position of interest, therefore $\nu=2\pi R - 2$. Thus, the threshold is also a function of separation, which varies from $8.1\;\sigma(R=1)$ at the inner working angle to $4.5\;\sigma(R=12.5)$ at $1^{\prime\prime}$.

The completeness, or sensitivity, of an observation is described by the true positive fraction \citep[TPF; see e.g.][]{Lafreniere2007,Wahhaj2013}. The signal level at a given completeness $S(R)$ is given by
\begin{equation}
\frac{S(R)}{\sigma(R)} = \frac{\tau(R)}{\sigma(R)} + C_\mathrm{st}^{-1}\left(\mathrm{TPF}|(2\pi R - 2)\right).
\end{equation}
The contrast associated with $S(R)$ at 95\% completeness (TPF=0.95) therefore varies from $10.1\;\sigma(R=1)$ at the inner working angle to $6.1\;\sigma(R=12.5)$ at $1^{\prime\prime}$. The contrast corresponding to one standard deviation $\sigma(R)$ was calculated using the \verb|contrast_curve| function in the VIP software package, which performs fake companion injection and retrieval to determine and compensate for signal losses owing to self-subtraction and over-subtraction effects \citep[see e.g.][]{Absil2013}. We injected planets in radial steps of one FWHM using a few planets at a time (depending on the frame size), with a spacing of 4-5 FWHM in between each planet. This was repeated until each separation was sampled along three directions in the image, evenly spaced in azimuth.

The resulting contrast limits are more conservative than the typically reported 50\% completeness (TPF=0.5) contour with $\tau(R) = 5\;\sigma(R)$, but can be easily traced back to a meaningful prediction for false positives in the image. A $5\;\sigma(R)$ contrast curve assumes a fixed FPF as a function of $R$ of $2.9\times10^{-7}$ and completeness of 50\%. In comparison, we have employed a radially-varying threshold that allows a higher number false positives (1\% versus $2.9\times10^{-7}\pi R^2_\mathrm{max}\approx0.01\%$ chance of a false positive within 1$^{\prime\prime}$) while accounting for the limited number of samples available in an annulus about the star at small angular separation. In addition, the upper limits on the contrast of point sources are obtained at a high value of completeness; that is, planets at the upper mass limit have 95\% probability of detection. 

% from Olivier
%using FPF levels that may be regarded as inappropriate owing to the number of independent samples in the region of interest (why use an FPF of 3e-7 if there are just 6 samples for instance?). Furthermore, standard contrast curves do not explicitly take into account the TPF (they implicitly rely on a 0.50 TPF, which is not very appropriate to set upper limits).
%- here we set a reasonable false alarm probability for the whole region of interest, and we spread it equally onto 12.5 annuli of 1 FWHM in width
%- we derive a detection threshold per annulus based on a Gaussian assumption, taking into account the small sample statistics. That threshold changes as a function of angular separation due to the changing number of samples in the annuli.
%- based on this detection threshold, we compute the contrast of a companion that would be detected 95% of the time (still using Gaussian assumption + small sample statistics)
%- we use this contrast to define our sensitivity limit in a conservative way (companions with contrast above this limit would be missed at most 5% of the time)

\clearpage

\bibliography{RuaneLibrary}

\clearpage

\end{document}